\documentclass[a4paper,11pt]{article}
\usepackage{pos}
\usepackage{slashed}
\usepackage{tikz}
\usetikzlibrary{shapes,arrows,positioning,automata,backgrounds,calc,er,patterns}
\usepackage{tikz-feynman}
\tikzfeynmanset{compat=1.1.0}
\usepackage{xcolor}
\usepackage{makecell}
\usepackage{rotating}
\usepackage{pdflscape}
\usepackage{float}
\newcommand{\Tr}[4]{\mathrm{Tr}(\slashed{#1}\slashed{#2} \slashed{#3} \slashed{#4})}
\newcommand{\ome}{(1-\epsilon)}
\newcommand{\omxi}{(1-x_i)}

\newcommand{\omxk}{(1-x_k)}

\newcommand{\fb}{b}
\newcommand{\fc}{\tilde{b}}

\newcommand{\Pqg}{P_{qg}}

\newcommand{\Pgg}{P_{gg}}

\newcommand\nameqgg{qgg \to q}
\newcommand\nameqpp{q\gamma\gamma \to q}

\newcommand{\Pqgg}{P_{\nameqgg}}
\newcommand{\Rqgg}{R_{\nameqgg}}
\newcommand{\Pqpp}{P_{\nameqpp}}
\newcommand{\Rqpp}{R_{\nameqpp}}

\title{Leading Order Triple Collinear Splitting Functions Revisited}
\author*[a]{Oscar Braun-White}

\affiliation[a]{Institute for Particle Physics Phenomenology,\\
Department of Physics,\\
Durham University, Durham, DH1 3LE, UK}

\emailAdd{oscar.r.braun-white@durham.ac.uk}

\abstract{I review the factorisation properties of tree level amplitudes when three particles $i$, $j$, $k$ are collinear.
The triple collinear splitting functions contain both iterated single unresolved contributions, and genuine double unresolved contributions. I make this explicit by rewriting the known triple collinear splitting functions for a quark and two gluons in terms of products of two-particle splitting functions, and a remainder that is explicitly finite when any two of $\{i,j,k\}$ are collinear.}

\FullConference{%
  Loops and Legs in Quantum Field Theory - LL2022,\\
  25-30 April, 2022\\
  Ettal, Germany
}

\begin{document}

\maketitle

\section{Introduction}

Theoretical predictions of QCD observables at the LHC need to be able to match the increasing levels of experimental precision. This mandates calculating corrections to the hard-partonic cross sections using perturbation theory in $\alpha_s$. QCD has been well-studied at next-to-leading order (NLO) in $\alpha_s$, including automated programs manipulating infrared subtraction terms~\cite{herwig:2015jjp,Sherpa:2019gpd,powheg:2010xd,madgraph:2011uj}. There are now also many results at next-to-next-to-leading order (NNLO) using NNLO subtraction schemes~\cite{TorresBobadilla:2020ekr}, such as~\cite{Currie:2016bfm,Czakon:2019tmo,Boughezal:2015dva,Gehrmann-DeRidder:2015wbt, Boughezal:2015ded,Campbell:2016lzl,Chen:2019zmr,Czakon:2015owf,Catani:2019iny,Chawdhry:2019bji,Chawdhry:2021hkp,Czakon:2021mjy}. Theoretical predictions at next-to-next-to-next-to-leading order (N3LO) are available for a few processes~\cite{Anastasiou:2016cez,Dulat:2018bfe,Duhr:2020seh,Chen:2021isd,Chen:2021vtu,Billis:2021ecs,Chen:2022cgv,Neumann:2022lft}. These results exploit the favourable $2 \to 1$ kinematics and there is currently no general N3LO subtraction scheme. The cancellation of infrared divergences across multi-loop corrections and multi-leg corrections is guaranteed by the KLN theorem at each order in $\alpha_s$. However, achieving the necessary analytic control of the corrections to balance against numerical calculations is extremely complex. The unresolved limits relevant for NLO are single unresolved, where one particle is soft or two are collinear. The unresolved limits at NNLO are the double unresolved limits of tree amplitudes~\cite{campbell,Catani:1998nv,Catani:1999ss,Kosower:2002su}, as well as the single unresolved limit of one-loop amplitudes~\cite{Bern:1994zx,Bern:1998sc,Kosower:1999rx,Bern:1999ry}. At N3LO, the relevant limits are the triple unresolved limits of tree amplitudes~\cite{Catani:2019nqv,DelDuca:1999iql,DelDuca:2019ggv,DelDuca:2020vst,DelDuca:2022noh}, the double unresolved limits of one-loop amplitudes~\cite{Catani:2003vu,Sborlini:2014mpa,Badger:2015cxa,Zhu:2020ftr,Catani:2021kcy,Czakon:2022fqi} and the single unresolved limits of two-loop amplitudes~\cite{Bern:2004cz,Badger:2004uk,Duhr:2014nda,Li:2013lsa,Duhr:2013msa}. These factorisation limits are also key in quantifying the accuracy of parton branching algorithms in event generators, and how these algorithms can eventually be extended to increase their logarithmic accuracy, see for example Refs.~\cite{Li:2016yez,Hoche:2017iem,Dulat:2018bfe,Dulat:2018vuy,Dasgupta:2020fwr,Loschner:2021keu,Gellersen:2021eci}. The work discussed in~\cite{Braun-White:2022rtg} exposes the single and double unresolved singularities hidden within leading order triple collinear splitting functions. These proceedings discuss two of these splitting functions. Revealing hidden structures at NNLO (and eventually at N3LO) is the key to simplifying NNLO subtraction schemes (and creating a viable N3LO subtraction scheme). 

\section{Triple Collinear Splitting Functions}

Time-like tree-level triple collinear splitting functions are extracted from colour-ordered sub-amplitudes, where the notation of Ref.~\cite{campbell} is used. For a colour-ordered sub-amplitude, 
\begin{equation}
    \mathcal{A} (...,i,j,k,...),
\end{equation}
particle $i$ is colour-connected to $j$, which is colour-connected to $k$. In the limit where $\{i,j,k\}$ become collinear, the triple collinear splitting function is extracted from the amplitude,
\begin{equation}
\label{eq:TC1}
	|\mathcal{A} (...,i,j,k,...)|^2 \rightarrow P_{abc \to P}(i,j,k) |\mathcal{A} (...,P,...)|^2 .
\end{equation}
Partons $i,j,k$ of particle type $a,b,c$ cluster to form particle $P$ with the sum of the momenta of $\{i,j,k\}$. 

The Lorentz invariant quantities,
\begin{equation}
	s_{i,\ldots,n} \equiv (p_{i}+...+p_{n})^2,
\end{equation}
are used, in terms of parton momenta. Using the massless quark limit suitable for high energy collisions and gluons being massless, $s_{ij} = 2p_i \cdot p_j = 2E_iE_j(1-\cos\theta_{ij})$, where $E_i$, $E_j$ are the energies of particles $i$, $j$ and $\theta_{ij}$ is the angle between them. Collinear and soft singularites can be expressed in terms of these Lorentz invariant quantities. $s_{ij}$ approaches zero if $i,j$ are collinear or at least one is soft. We work in $d=4-2\epsilon$ dimensions. The triple collinear limit is the region where $s_{ij},$ $s_{jk}$, $s_{ik}$, $s_{ijk}$ become small. We can write $p_i = x_i p_P$, $p_j = x_j p_P$ and $p_k = x_k p_P$ with $x_i+x_j+x_k = 1$.  In practice, a spectator momentum $\ell$ is used to define the momentum fractions, $s_{i\ell}= x_i s_{P\ell}$. In these proceedings we will focus only on two splittings which illustrate all the features exposed in~\cite{Braun-White:2022rtg}:  $\nameqpp$, $\nameqgg$. We are using a shorthand notation for the splitting functions,
\begin{equation}
P_{abc\to P} (i,j,k) \equiv P_{abc \to P}(x_i, x_j, x_k; s_{ij},s_{ik},s_{jk},s_{ijk}).
\end{equation} 

\section{Singularity structure of the triple collinear splitting function} \label{sec:singstructure}

Can you decompose triple collinear splitting functions into a strongly-ordered iterated collinear splitting and a remainder which is finite when any two of $\{i,j,k\}$ are collinear?

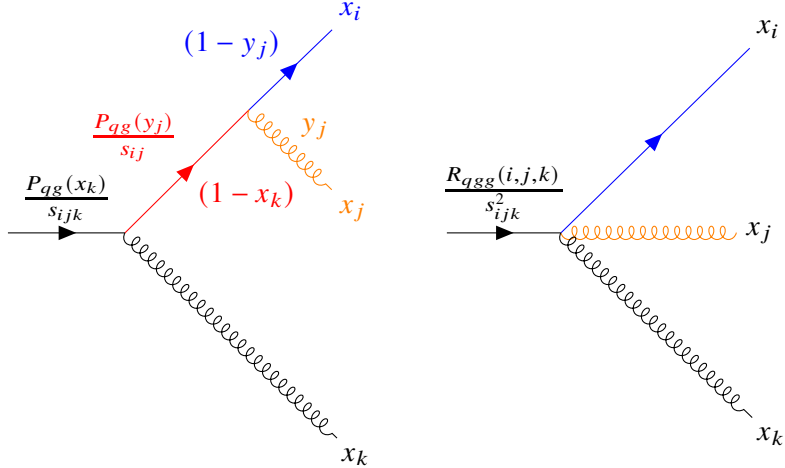
\begin{figure}[th!]
    \centering
\begin{tikzpicture} 
\begin{feynman}
\vertex (a); 
\vertex [right=4em of a] (b);
\vertex [above right=6em of b]  (d) ; 
\vertex [above right=of d]  (f1) {\(\color{blue}x_i\)};
\vertex [below right=of d]  (f2) {\(\color{orange} x_j\)};
\vertex [below right=10em of b]  (f3) {\( \color{black} x_k\)};

\vertex [right=15 em of a] (aa);
\vertex [right=of aa] (bb);
\vertex [above right=9em of bb]  (ff1) {\(x_i\)};
\vertex [right=6em of bb]  (ff2) {\(x_j\)};
\vertex [below right=9em of bb]  (ff3) {\(x_k\)};

\diagram* {
(a) -- [fermion, edge label=\( \frac{P_{qg} (x_k)}{s_{ijk}} \)] (b),
(b) -- [fermion, edge label=\(\frac{P_{qg}(y_j)}{s_{ij}} \), edge label'=\( (1-x_k) \), color={red}] (d), 
(d) -- [fermion, edge label=\( (1-y_j) \),color={blue}] (f1), 
(d) -- [gluon, edge label=\(y_j\),color={orange}] (f2),
(b) -- [gluon,color={black}] (f3), 
(aa) -- [fermion, edge label=\( \frac{R_{qgg}(i,j,k)}{s_{ijk}^2} \) ] (bb),
(bb) -- [fermion,color={blue}] (ff1),
(bb) -- [gluon,color={orange}] (ff2),
(bb) -- [gluon,color={black}] (ff3), 
};
\end{feynman}
\end{tikzpicture}    

    \caption{On the left: An iterated single-collinear contribution to the triple collinear splitting function. On the right: A remainder function $R_{abc\to P}$ which contains the parts of the triple collinear splitting function that are not contained in the strongly-ordered, iterated contributions. }
    \label{fig:iterated}
\end{figure}

This can be illustrated in the left of ~Fig.~\ref{fig:iterated}, where $P$ undergoes a strongly-ordered iterated splitting into $i,j,k$ and the right of ~Fig.~\ref{fig:iterated}, where $P$ undergoes a direct $1\rightarrow 3$ splitting. The momentum fraction of the second iterated splitting $y_j$ can be related to $\{x_i,x_j,x_k\}$ by momentum conservation: $ x_j = (1-x_k) y_j$. The invariants in the denominator of the $1\rightarrow 2$ splitting functions are $s_{ijk}$ and $s_{ij}$ because these are the propagator contributions to the amplitudes. 

In other words, we rewrite the triple collinear splitting functions as
\begin{equation}
\label{eq:Psplit}
P_{abc\to P} (i,j,k) = \sum_{{\mathrm perms }} 
\frac{1}{s_{ijk}} P_{(ab)c \to P}\left(x_k\right)
\frac{1}{s_{ij}} P_{ab \to (ab)}\left(\frac{x_j}{1-x_k}\right) 
+ \frac{1}{s_{ijk}^2}R_{abc\to P} (i,j,k)
\end{equation}
where $P_{ab \to (ab)}$ are the usual spin-averaged two-particle splitting functions and the remainder $R_{abc\to P} (i,j,k)$ depends on the momentum fractions and small invariants. 

\section{New Basis}

In order to decide which terms in $P_{abc \to P} (i,j,k)$ should be in the remainder $R_{abc \to P} (i,j,k)$, we express the splitting functions in terms of a different basis which organises how terms contribute in single collinear limits. The new basis is constructed out of traces of four gamma matrices contracted with momenta:
\begin{equation}
    \Tr{i}{j}{k}{\ell} = s_{ij} s_{kl} + s_{il} s_{jk} - s_{ik}s_{jl}.
\end{equation}
The useful property of these traces is that $\Tr{i}{j}{k}{\ell}/s_{ij}$ is not singular in the $i,j$ collinear limit. This be seen by expanding the trace in the $i,j$ collinear limit, where $Y_{ij} \equiv s_{il} s_{jk} - s_{ik} s_{jl} \to \mathcal{O} (\sqrt{s_{ij}})$. Similarly $\Tr{i}{j}{k}{\ell}/s_{jk}$, $\Tr{i}{j}{k}{\ell}/s_{kl}$ and $\Tr{i}{j}{k}{\ell}/s_{il}$ are not singular in the $(j,k)$, $(k,l)$, $(i,l)$ collinear limits respectively. While the same is not true for $\Tr{i}{j}{k}{\ell}/s_{ik}$ or $\Tr{i}{j}{k}{\ell}/s_{jl}$. When we work in the $i,j,k$ collinear limit, we allow the momentum of a spectator $\ell$ to be normalised such that, $\Tr{i}{j}{k}{\ell} = x_k s_{ij} + x_i s_{jk} - x_j s_{ik}$. The new basis is described in full in~\cite{Braun-White:2022rtg}.

We introduce 
\begin{equation}
\label{eq:Wdef}
        W_{ij} = (x_i s_{jk} - x_j s_{ik})^2 - 
        \frac{ 2} {\ome}
        \frac{ x_i x_j x_k} {(1-x_k)}
        s_{ij} s_{ijk},
\end{equation}
such that terms like $W_{ij}/s_{ij}^2$ in the new basis do not contribute in the $i,j$ collinear limit and are $\mathcal{O} (1/\sqrt{s_{ij}})$. The second term removes the $i,j$ collinear contribution from the first term. This is an integrable singularity which vanishes upon azimuthal integration in $d$ dimensions, where we take the azimuthal angle with respect to the $(ij)$ direction. Using Ref.~\cite{Dulat:2018vuy}, we can write
\begin{equation}
\label{eq:azim}
(x_i s_{jk} - x_j s_{ik})^2 = \frac{4 x_i x_j x_k}{\omxk} s_{ij} s_{ijk} \cos^2 \phi_{ij,kl},
\end{equation}
such that $W_{ij}$ has the form,
\begin{equation}
    W_{ij} = \frac{4 x_i x_j x_k}{\omxk} s_{ij} s_{ijk} \left( \cos^2 \phi_{ij,kl} - \frac{1}{2 \ome} \right).
\end{equation}

We also introduce the auxiliary functions, 
\begin{equation}
\label{eq:A0def}
    A_0(x,y) = 1 - \frac{(1-x)}{(1-y)}, \text{  } B_0(x,y) = 1 + \frac{2x(x-2)}{(1-y)^2} + \frac{4x}{(1-y)}. 
\end{equation}

\section{Results}

General features of the results (for all QCD triple collinear splitting functions) are as follows: All terms with two of $\{i,j,k\}$ collinear, appear as a permutation of $\frac{1}{s_{ijk}} P_{(ab)c \to P}\left(x_k\right)
\frac{1}{s_{ij}} P_{ab \to (ab)}\left(\frac{x_j}{1-x_k}\right) $, or $P \times P$ for short. In practice, due to colour-ordering, many coefficients in the new basis are zero. All splitting functions are describable with one trace ordering (appropriate to the colour-ordering) if $\{\gamma_\mu, \gamma_\nu\}$ is used.
The full set of results are examined in~\cite{Braun-White:2022rtg}.

We expose all single and double unresolved singularities contained within the triple collinear splitting functions and introduce the concept of internal and external singularities when in a triple collinear limit. Internal singularities in a $i,j,k$ collinear limit involve only small invariants in $\{s_{ij},s_{jk},s_{ik},s_{ijk}\}$.
\begin{itemize}
    \item {\bf Internal} single collinear singularities like $1/s_{ij}$ appear only in $P \times P$ terms (the iterated two-particle splitting contributions). 
    \item {\bf Internal} single soft singularities like $s_{ik}/s_{ij}/s_{jk}$ appear only in $R_{abc \to P}$.
\end{itemize}
External singularities involve singularities in momentum fractions $x_I$, with reference to a spectator particle and are inherited from the parent multi-leg amplitudes before taking the $i,j,k$ collinear limit. 
\begin{itemize}
    \item When {\bf external} single collinear singularities like $1/x_i$ appear in $P_{abc \to P}$, they are all contained in $P \times P$ terms.
    \item When {\bf external} single collinear singularities like $1/x_i$ {\bf do not} appear in $P_{abc \to P}$, there could be terms proportional to $1/x_i$ in $P \times P$ and $R_{abc \to P}$ which cancel. 
    \item {\bf External} single soft singularities like $x_i/x_j/s_{ij}$ appear only in the iterated  $P \times P$ terms.
\end{itemize}
Double soft singularities are defined as external because they require reference to some spectator.

\subsection{Two gluons with a collinear quark or antiquark}

\vspace{3mm}\noindent (a)  
In the case where the gluons are abelianised ($\tilde{g}$) or two photons are collinear to the quark, the splitting function is symmetric under the exchange of the two bosons ($j,k$).  We find,
\begin{eqnarray}
\label{eq:Pqpp}
\Pqpp(i,j,k) &=& 
		\frac{\Pqg(x_k)}{s_{ijk}}  
		\frac{\Pqg\left(\frac{x_j}{1-x_k}\right)}{s_{ij}} 
	+	\frac{\Pqg(x_j)}{s_{ijk}}  
		\frac{\Pqg\left(\frac{x_k}{1-x_j}\right)}{s_{ik}} \nonumber \\
&&		+ \frac{1}{s_{ijk}^2} \Rqpp (i,j,k),
\end{eqnarray}
where
\begin{equation}
\label{eq:Rqpp}
\Rqpp (i,j,k) =
  \fc_0^\text{sub}(x_i,x_j,x_k) -  \frac{\ome^2}{(1-x_k)} \frac{\Tr{j}{i}{k}{\ell}}{s_{ij}} + \fc^\text{sub}(x_i,x_j,x_k)	\frac{s_{ijk} \Tr{j}{i}{k}{\ell}}{s_{ij}s_{ik}} + (j\leftrightarrow k),
\end{equation}
and
\begin{eqnarray}
\label{eq:fc0}
    \fc_0^\text{sub}(x_i,x_j,x_k) &=& \ome\left(1 - \ome A_0(x_j,x_k) \right), \\
\label{eq:fc}
	\fc^\text{sub}(x_i,x_j,x_k) &=& - \frac{x_k \Pqg(x_k)}{x_j (1-x_i)} + \frac{2}{(1-x_i)} - 2 \ome + \frac{1}{2} \ome^2.
\end{eqnarray}
Eqs.~\eqref{eq:Pqpp}--\eqref{eq:fc} are equivalent to Eq.~(5.6) in Ref.~\cite{campbell} up to a normalisation of a factor of 4.

The behaviour of the $\Pqpp$ triple collinear splitting function in the limit where individual momentum fractions are small is in Table~\ref{table:qpp}. There is no singular behaviour as $x_i \to 0$ reflecting the fact that there is no singularity when the quark and spectator momentum are collinear and that there is no soft quark singularity.
We also see that there are contributions from both the strongly-ordered contribution and from $\Rqpp$ when $x_j \to 0$ and $x_k \to 0$ that do not cancel in the full $\Pqpp$ splitting function. 
However, only the strongly-ordered term contributes in the soft $j$ or soft $k$ limits, 
\begin{eqnarray}
    \Pqpp(i,j,k) &\stackrel{j~{\rm soft}}{\longrightarrow}& \frac{2x_i}{s_{ij}x_j} \frac{1}{s_{ik}}\Pqg(x_k),\\
   \Pqpp(i,j,k) &\stackrel{k~{\rm soft}}{\longrightarrow}& \frac{2x_i}{s_{ik}x_k} \frac{1}{s_{ij}}\Pqg(x_j).   
\end{eqnarray}
It can be seen that the strongly-ordered terms contribute the full double soft $j,k$ limit (a product of two eikonal factors) and there are no contributions from $\fc (x_i,x_j,x_k)$. There are no other double soft singularities.

%\begin{landscape}
%\begin{sidewaystable}[p]
\begin{table}[H]
\centering
%\footnotesize
\begin{center}
\begin{tabular}{|c|| c |c|| c|} 
 \hline
 $\nameqpp$ & 
 \makecell[c]{ $P \times P $ terms}
 & $\frac{1}{s_{ijk}^2} \Rqpp(i,j,k)$ 
 & $\frac{1}{s_{ijk}^2} \Pqpp(i,j,k)$ \\  
 \hline\hline
 $x_i \rightarrow 0$ 
 &\makecell[c]{0}
 & \makecell[c]{0}
 & \makecell[c]{0} \\ 
 \hline
 $x_j \rightarrow 0$ 
 & \makecell[l]{  
 $\phantom{\bigg[}$ \\ 
 $+\frac{1}{s_{ij}s_{ijk}} \frac{ x_i}{x_j} \bigg[2 \Pqg(x_k) \bigg] $ \\
 $+ \frac{1}{s_{ik}s_{ijk}} \frac{1}{x_j} \bigg[2 \Pqg(x_k) \bigg]$ } 
 & \makecell[l]{ $+ \frac{1}{s_{ij}s_{ik}} \frac{x_i}{x_j} \bigg[\Pqg(x_k) \bigg]$ \\ 
 $+ \frac{1}{s_{ij}s_{ijk}} \frac{x_i}{x_j} \bigg[-\Pqg(x_k) \bigg]$ \\
 $+ \frac{1}{s_{ik}s_{ijk}} \frac{1}{x_j} \bigg[-\Pqg(x_k) \bigg]$} 
 & \makecell[l]{ $+ \frac{1}{s_{ij}s_{ik}} \frac{x_i}{x_j} \bigg[\Pqg(x_k) \bigg]$ \\ 
 $+ \frac{1}{s_{ij}s_{ijk}} \frac{x_i}{x_j} \bigg[\Pqg(x_k) \bigg]$ \\
 $+ \frac{1}{s_{ik}s_{ijk}} \frac{1}{x_j} \bigg[\Pqg(x_k) \bigg]$} \\
 \hline
 $x_k \rightarrow 0$ 
 & \makecell[l]{   $\phantom{\bigg[}$ \\ 
 $+\frac{1}{s_{ij}s_{ijk}} \frac{ 1}{x_k} \bigg[2 \Pqg(x_j) \bigg] $ \\
 $+ \frac{1}{s_{ik}s_{ijk}} \frac{x_i}{x_k} \bigg[2 \Pqg(x_j) \bigg]$ } 
 & \makecell[l]{ $+ \frac{1}{s_{ij}s_{ik}} \frac{x_i}{x_k} \bigg[\Pqg(x_j) \bigg]$ \\ 
 $+ \frac{1}{s_{ij}s_{ijk}} \frac{1}{x_k} \bigg[-\Pqg(x_j) \bigg]$ \\
 $+ \frac{1}{s_{ik}s_{ijk}} \frac{x_i}{x_k} \bigg[-\Pqg(x_j) \bigg]$} 
 & \makecell[l]{$+ \frac{1}{s_{ij}s_{ik}} \frac{x_i}{x_k} \bigg[\Pqg(x_j) \bigg]$ \\
  $+ \frac{1}{s_{ij}s_{ijk}} \frac{1}{x_k} \bigg[\Pqg(x_j) \bigg]$ \\
  $+ \frac{1}{s_{ik}s_{ijk}} \frac{x_i}{x_k} \bigg[\Pqg(x_j) \bigg]$} \\
 \hline
\end{tabular}
\end{center}
\caption{Singular behaviour of the $\Pqpp$ triple collinear splitting function in the limit where individual momentum fractions are small. The contributions from the iterated two-particle splittings are shown in column 2, while the contributions from $\Rqpp$ are shown in column 3 and the contributions for the entire splitting function $\Pqpp$ are shown in column 4. Each row shows the singular limit for a different momentum fraction tending to zero. The vertical displacement within each cell is organised by $\{s_{ij},s_{jk},s_{ik},s_{ijk}\}$.}
\label{table:qpp}
\end{table}
%\end{sidewaystable}
%\end{landscape}

\vspace{3mm}\noindent (b) 
In the case where gluon $j$ is colour-connected to quark $i$ and gluon $k$, we find that,
\begin{eqnarray}
\label{eq:Pqgg}
\Pqgg(i,j,k) &=& 
\frac{\Pqg(x_k)}{s_{ijk}} \frac{\Pqg\left(\frac{x_j}{1-x_k}\right)}{s_{ij}}
+
\frac{\Pqg(1-x_i)}{s_{ijk}}   \frac{\Pgg\left( \frac{x_j}{1-x_i} \right)}{s_{jk}} 
\nonumber \\
&&+\frac{1}{s_{ijk}^2} \Rqgg(i,j,k),
\end{eqnarray}
where
\begin{eqnarray}
\label{eq:Rqgg}
\Rqgg (i,j,k) &=& 
\frac{2 \ome}{(1-x_i)^2}  \frac{W_{jk}}{s_{jk}^2} 
+ \frac{4 \ome x_k }{(1-x_i)^{2}} \frac{\Tr{i}{j}{k}{\ell}}{s_{jk}} + \frac{\ome^2}{(1-x_k)}  \frac{\Tr{i}{j}{k}{\ell}}{s_{ij}}
\nonumber \\
&&
+ \fb_0 (x_i,x_j,x_k) 
+  \fb(x_i,x_j,x_k) \frac{s_{ijk} \Tr{i}{j}{k}{\ell}}{s_{ij}s_{jk}}, 
\end{eqnarray}
and
\begin{eqnarray}
\label{eq:fb0}
\fb_0 (x_i,x_j,x_k) &=& 
\ome \left(B_0(x_k,x_i) -1 + \ome A_0(x_i,x_k)\right),\\
\label{eq:fb}
\fb(x_i,x_j,x_k) &=&- \frac{x_j \Pqg(x_j)}{x_k (1-x_i)} - \frac{2 x_k \Pqg(x_k)}{x_j (1-x_i)} + \frac{4}{(1-x_i)} - 3 \ome.
\end{eqnarray}
Eqs.~\eqref{eq:Pqgg}--\eqref{eq:fb} are equivalent to Eq.~(5.5) in Ref.~\cite{campbell} up to a normalisation of a factor of 4. 

We observe that $\fb$ contains inverse powers of $x_j$ and $x_k$. The behaviour of the $\Pqgg$ triple collinear splitting function in the limit where individual momentum fractions are small is in Table~\ref{table:qgg}. We see that there is no singular behaviour as $x_i \to 0$.  This is because there is no singularity when the quark and spectator are collinear and there is no soft quark singularity.  When $x_j \to 0$, we see contributions from both the strongly-ordered contribution and from $\Rqgg$ which cancel in the full $\Pqgg$ splitting function.
When $x_k \to 0$, we see that the contributions from the strongly-ordered contribution and from $\Rqgg$ do not cancel in the full $\Pqgg$ splitting function.

%\begin{landscape}
%\begin{sidewaystable}[p]
\begin{table}[t]
\centering
%\footnotesize
\begin{center}
\begin{tabular}{|c|| c |c|| c|} 
 \hline
 $\nameqgg$ & \makecell[c]{
  $P \times P $ terms
 } 
 & $\frac{1}{s_{ijk}^2} \Rqgg(i,j,k)$ 
 & $\frac{1}{s_{ijk}^2} \Pqgg(i,j,k)$ \\  
 \hline\hline
 $x_i \rightarrow 0$ 
 &\makecell[c]{0}
 & \makecell[c]{0}
 & \makecell[c]{0} \\ 
 \hline
 $x_j \rightarrow 0$ 
 & \makecell[l]{  \\  $+ \frac{1}{s_{ij}s_{ijk}} \frac{ x_i}{x_j} \bigg[ 2 \Pqg(x_k) \bigg]  $  \\ $+ \frac{1}{s_{jk}s_{ijk}} \frac{ x_k}{x_j} \bigg[ 2 \Pqg(x_k) \bigg]$} 
 & \makecell[l]{  \\  $+ \frac{1}{s_{ij}s_{ijk}} \frac{ x_i}{x_j} \bigg[ -2 \Pqg(x_k) \bigg]  $  \\ $+ \frac{1}{s_{jk}s_{ijk}} \frac{ x_k}{x_j} \bigg[ -2 \Pqg(x_k) \bigg]$} 
 & \makecell[c]{  \\ 0 \\} \\
 \hline
 $x_k \rightarrow 0$ 
 & \makecell[l]{   \\ $+\frac{1}{s_{ij}s_{ijk}} \frac{ 1}{x_k} \bigg[2 \Pqg(x_j) \bigg] $ \\$+ \frac{1}{s_{jk}s_{ijk}} \frac{x_j}{x_k} \bigg[2 \Pqg(x_j) \bigg]$ } 
 & \makecell[l]{ $+ \frac{1}{s_{ij}s_{jk}} \frac{x_j}{x_k} \bigg[ \Pqg(x_j) \bigg] $ \\ $+\frac{1}{s_{ij}s_{ijk}} \frac{ 1}{x_k} \bigg[- \Pqg(x_j) \bigg] $ \\$+ \frac{1}{s_{jk}s_{ijk}} \frac{x_j}{x_k} \bigg[-\Pqg(x_j) \bigg]$} 
 & \makecell[l]{$+ \frac{1}{s_{ij}s_{jk}} \frac{x_j}{x_k} \bigg[ \Pqg(x_j) \bigg] $ \\ $+\frac{1}{s_{ij}s_{ijk}} \frac{ 1}{x_k} \bigg[\Pqg(x_j) \bigg] $   \\$+ \frac{1}{s_{jk}s_{ijk}} \frac{x_j}{x_k} \bigg[\Pqg(x_j) \bigg]$} \\
 \hline
\end{tabular}
\end{center}
\caption{Singular behaviour of $\Pqgg$ in the limit where individual momentum fractions are small.}
\label{table:qgg}
\end{table}
%\end{sidewaystable}
%\end{landscape}

In the soft $k$ limit, only the strongly-ordered term contributes and we recover the expected limit describing collinear partons $i$ and $j$ with the soft gluon $k$ radiated between the colour-connected partners $j$ and $\ell$,
\begin{equation}
    \Pqgg(i,j,k) \stackrel{k~{\rm soft}}{\longrightarrow} \frac{2x_j}{s_{jk}x_k} \frac{1}{s_{ij}}\Pqg(x_j).
\end{equation}

However, in the soft $j$ limit the $1/x_j/s_{ij}$ and $1/x_j/s_{jk}$ terms cancel between the $P \times P$ and $\Rqgg$ contributions, such that
\begin{eqnarray}
    \frac{1}{s_{ijk}^2} \Rqgg(i,j,k) &
    \stackrel{j~{\rm soft}}{\longrightarrow}& \left(
    -\frac{2x_i}{x_j s_{ij}} 
    -\frac{2x_k}{x_j s_{jk}} 
    +\frac{2s_{ik}}{s_{ij}s_{jk}}
    \right)
    \frac{1}{s_{ik}}\Pqg(x_k),\\    
    \Pqgg(i,j,k) &
    \stackrel{j~{\rm soft}}{\longrightarrow}& \frac{2 s_{ik}}{s_{ij}s_{jk}}\frac{1}{s_{ik}}\Pqg(x_k).
\end{eqnarray}
This is precisely as expected for the emission of a soft gluon between the hard (and collinear) radiators $i$ and $k$. There are also double soft singularities when gluons $j,k$ are soft. These are contained iteratively in the $P \times P$ contributions and in $\Rqgg (i,j,k)$, 
\begin{eqnarray} 
\label{eq:DsoftinPxPqgg}
\frac{\Pqg(1-x_i)}{s_{ijk}}  \frac{\Pgg\left(\frac{x_j}{1-x_i} \right)}{s_{jk}} && + \frac{\Pqg(x_k)}{s_{ijk}}  \frac{\Pqg\left(\frac{x_j}{1-x_k} \right)}{s_{ij}}\nonumber \\
&&\stackrel{j,k~{\rm soft}}{\longrightarrow}
\frac{2}{\omxi s_{ijk} s_{jk}} \Pgg\left(\frac{x_j}{1-x_i}\right) + \frac{4}{x_j x_k s_{ijk} s_{ij}},\\
\label{eq:DsoftinRqgg}
\frac{1}{s_{ijk}^2}
\Rqgg(i,j,k) &&\stackrel{j,k~{\rm soft}}{\longrightarrow}
 \frac{2 \ome W_{jk}}{(1-x_i)^2 s_{jk}^2 s_{ijk}^2}
 - \left( 
\frac{2}{x_k (1-x_i)} + \frac{4}{x_j\omxi}
 \right) \frac{\Tr{i}{j}{k}{\ell}}{s_{ij} s_{jk} s_{ijk}}. \nonumber \\
\end{eqnarray}

\section{Conclusions}

In these proceedings, we have rewritten $\Pqpp$ and $\Pqgg$ to expose the single and double unresolved limits. These two splitting functions, when decomposed, display all the features which are present in the other QCD triple collinear splitting functions~\cite{Braun-White:2022rtg}. We have isolated the strongly-ordered iterated contributions as products of the usual spin-averaged two-particle splitting functions (generically $P \times P$) and a remainder function $R_{abc\to P} (i,j,k)$ that is finite when any pair of $\{i,j,k\}$ are collinear. This work allows us to appreciate the interplay between NLO and NNLO, which is important for the development of efficient infrared subtraction schemes. Similar work on the quadruple collinear limit could be useful for progressing to N3LO. 

\begin{acknowledgments}
{I thank Aude Gehrmann-De Ridder, Thomas Gehrmann and Christian Preuss for useful discussions and of course my supervisor Nigel Glover.}
\end{acknowledgments}

\bibliographystyle{JHEP}
\bibliography{biblio}

\end{document}